\documentclass{article}
\pdfoutput=1 

\usepackage[utf8]{inputenc}
\usepackage{todonotes}
\usepackage{authblk}

\newcommand{\system}{{\sc KGPS}}
\newcommand{\knpl}{{\sc KGPS}}
\newcommand{\knps}{{\sc KGPS}}
\newcommand{\fullknps}{{\bf Knowledge Graph Programming System}}
\usepackage{graphicx}

\usepackage{hyperref}
\hypersetup{
    colorlinks=true,
    linkcolor=blue,
    filecolor=magenta,      
    urlcolor=cyan,
}

\title{Technical Report on Data Integration and Preparation}

\author[1]{El Kindi Rezig}
\author[1]{Michael Cafarella}
\author[2]{Vijay Gadepally}
\affil[1]{Computer Science and Artificial Intelligence Laboratory, MIT}
\affil[2]{Lincoln Laboratory, MIT}

\date{}
\begin{document}

\maketitle

\newpage
\begin{abstract}

AI application developers typically begin with a dataset of interest and a vision of the end analytic or insight they wish to gain from the data at hand. Although these are two very important components of an AI workflow, one often spends the first few weeks (sometimes months) in the phase we refer to as data conditioning. This step typically includes tasks such as figuring out how to prepare data for analytics, dealing with inconsistencies in the dataset, and determining which algorithm (or set of algorithms) will be best suited for the application. Larger, faster, and messier datasets such as those from Internet of Things sensors, medical devices or autonomous vehicles only amplify these issues. These challenges, often referred to as the three Vs (volume, velocity, variety) of Big Data, require low-level tools for data management, preparation and integration. In most applications, data can come from structured and/or unstructured sources and often includes inconsistencies, formatting differences, and a lack of ground-truth labels.

In this report, we highlight a number of tools that can be used to simplify data integration and preparation steps. Specifically, we focus on data integration tools and techniques, a deep dive into an exemplar data integration tool, and a deep-dive in the evolving field of knowledge graphs. Finally, we provide readers with a list of practical steps and considerations that they can use to simplify the data integration challenge. The goal of this report is to provide readers with a view of state-of-the-art as well as practical tips that can be used by data creators that make data integration more seamless. 

\end{abstract}
\newpage
\tableofcontents
\newpage

\section{Executive Summary}

Many AI application developers typically begin with a dataset of interest and a vision of the end analytic or insight they wish to gain from the data at hand. Although these are two very important components of the AI pipeline, one often spends the first few weeks (sometimes months) in the phase we refer to as data conditioning. This step typically includes tasks such as figuring out how to store data, dealing with inconsistencies in the dataset, and determining which algorithm (or set of algorithms) will be best suited for the application. Larger, faster, and messier datasets such as those from Internet of Things sensors \cite{sowe2014managing}, medical devices \cite{johnson2016mimic,saeed2011multiparameter} or autonomous vehicles \cite{gadepally2013framework} only amplify these issues. These challenges, often referred to as the three Vs (volume, velocity, variety) of Big Data, require low-level tools for data management and data cleaning/pre-processing. In most applications, data can come from structured and/or unstructured sources and often includes inconsistencies, formatting differences, and a lack of ground-truth labels. In practice, there is a wide diversity in the quality, readiness and usability of data \cite{lawrence2017data}.

It is widely reported that data scientists spend at least 80\% of their time doing data integration and preparation~\cite{Deng17}. This process involves finding relevant datasets for a particular data science task (i.e., data discovery~\cite{aurum}). After the data has been collected, it needs to be cleaned so that it can be used by subsequent ML models~\cite{dc2}. Data cleaning or data preparation refer to several data transformations to make the data reliable for future analysis. Example transformations include: (1)~finding and fixing errors in the data using rules (e.g., an employee salary should not exceed \$1M); (2)~normalizing value representations~\cite{pkduck}; (3)~imputing missing values~\cite{imputedb, fahes}; and (4)~detecting and reconciling duplicates. 

This document provides a survey of the state-of-the-art in data preparation and integration. To help make concepts clearer, we also discuss a tool developed by our team called Data Civilizer ~\cite{dc2}. Additionally, we detail the concept of Knowledge Graphs - a technique that can be used to describe structured data about a topic as a tool for organizing data with applications in data preparation, integration and beyond. Finally, this document highlights a number of practical steps, based largely on experience, that can be used to simplify data preparation and integration. A summary of these steps is given below:

\begin{enumerate}
    \item Data Organization 
    \begin{itemize}
        \item Leverage tools for schema integration
        \item Use standardized file formats and naming conventions
    \end{itemize}
    \item Data Quality and Discovery
    \begin{itemize}
        \item Maintain data integrity through enforceable constraints
        \item Leverage workflow management systems for testing various hypothesis on the data
        \item Include data version control
        \item Leverage data discovery tools
    \end{itemize}
    \item Data Privacy
    \begin{itemize}
        \item Leverage encrypted or secure data management systems as appropriate
        \item Limit data ownership
        \item Avoid inadvertent data releases
        \item Be aware that data aggregated may be sensitive
    \end{itemize}
    \item Infrastructure, Technology and Policy
    \begin{itemize}
        \item Use data lakes as appropriate
        \item Databases and files can be used together
        \item Leverage federated data access
        \item Access to the right talent
        \item Pay attention to technology selection, software licensing and software/hardware platforms
        \item Maintain an acquisition and development environment conducive to incorporating new technology advances
    \end{itemize}
\end{enumerate}
\newpage

\section{Data Integration Introduction and Challenges}

Many AI application developers typically begin with a dataset of interest and a vision of the end analytic or insight they wish to gain from the data at hand. Although these are two very important components of the AI pipeline, one often spends the first few weeks (sometimes months) in the phase we refer to as data conditioning. This step typically includes tasks such as figuring out how to store data, dealing with inconsistencies in the dataset, and determining which algorithm (or set of algorithms) will be best suited for the application. Larger, faster, and messier datasets such as those from Internet of Things sensors \cite{sowe2014managing}, medical devices \cite{johnson2016mimic,saeed2011multiparameter} or autonomous vehicles \cite{gadepally2011driver} only amplify these issues. These challenges, often referred to as the three Vs (volume, velocity, variety) of Big Data, require low-level tools for data management, data integration \cite{lenzerini2002data} and data preparation (this includes data cleaning/pre-processing). In most applications, data can come from structured and/or unstructured sources and often includes inconsistencies, formatting differences, and a lack of ground-truth labels. One interesting way to think about data preparedness is through the concept of data readiness levels as outlined in \cite{lawrence2017data} which is similar to the idea of technology readiness levels (TRL) \cite{sauser2006trl}. 

It is widely reported that data scientists spend at least 80\% of their time doing data integration and preparation~\cite{Deng17,stonebraker2018data}. This process involves finding relevant datasets for a particular data science task (i.e., data discovery~\cite{aurum}). After the data has been collected, it needs to be cleaned so that it can be used by subsequent ML models~\cite{dc2}. Data cleaning or data preparation refer to several data transformations to make the data reliable for future analysis. Example transformations include: (1)~finding and fixing errors in the data using rules (e.g., an employee salary should not exceed \$1M); (2)~normalizing value representations~\cite{pkduck}; (3)~imputing missing values~\cite{imputedb, fahes}; and (4)~detecting and reconciling duplicates. 

At a high level, the concept of data integration and preparation is the effort required to go from raw sensor data to information that can be used in further processing steps (see the Artificial Intelligence architecture provided in \cite{gadepally2019ai}). Sometimes this phase is also referred to as data wrangling. Typically, each of these data integration and preparation tasks can be cumbersome, require significant domain knowledge, and represent a significant hurdle in developing an end-to-end application. Many of the recent algorithmic advances have, in fact, occurred in areas where "prepared" data can be found. For example, advances in image classification were largely driven by the availability of the ImageNet dataset \cite{deng2009imagenet}, advances in handwriting recognition by the Modified National Institute of Standards and Technology (MNIST) dataset \cite{xiao2017fashion}, and advances in video recognition by the Moments in Time dataset \cite{monfort2018moments}. Other popular datasets such as CIFAR-10 \cite{krizhevsky2010convolutional}, and Internet packet capture traces \cite{gadepally2018hyperscaling,fontugne2010mawilab,do2020classifying} have played important roles in advancing certain classes of algorithms and genres of applications. Data integration and preparation is often a first step in getting one's data into such a form.

There are a number of research efforts and organizations aiming to reduce the data integration and preparation barrier to entry. For example, there are techniques to do spectral fingerprinting of datasets to look for outliers \cite{gadepally2014big}, techniques for modelling the background of datasets \cite{gadepally2015using}, and techniques for organizing data into federated data management systems \cite{tan2017enabling,hausenblas2013apache, gadepally2016bigdawg, halperin2014demonstration}. As examples of end-to-end data applications,  \cite{mattson2017demonstrating} describes an oceanographic data use-case  and \cite{elmore2015demonstration} describes a medical use-case.

In this report, we highlight a number of tools that can be used to simplify data integration and preparation steps. Specifically, we focus on data integration tools and techniques, a deep dive into an exemplar data integration tool, and a deep-dive in the evolving field of knowledge graphs. Finally, we provide readers with a list of practical steps they can use to simplify the data integration challenge. The goal of this report is to provide readers with a view of state-of-the-art as well as practical tips that can be used by data creators that make data integration more seamless. 

\newpage

\section{Tools and Techniques}

\subsection{Data Integration Challenges}
%talk more about the problems
%related work: dataxformer, tamer
It is widely reported that data scientists spend at least 80\% of their time doing data integration and preparation~\cite{Deng17}. This process involves finding relevant datasets for a particular data science task (i.e., data discovery~\cite{aurum}). After the data has been collected, it needs to be cleaned so that it can be used by subsequent ML models~\cite{dc2}. Data cleaning or data preparation refer to several data transformations to make the data reliable for future analysis. Example transformations include: (1)~finding and fixing errors in the data using rules (e.g., an employee salary should not exceed \$1M); (2)~normalizing value representations~\cite{pkduck}; (3)~imputing missing values~\cite{imputedb, fahes}; and (4)~detecting and reconciling duplicates,

\subsection{Data Integration Tasks}

\begin{figure}[h!]
\centering
\includegraphics[width=10cm,bb=0 0 700 500]{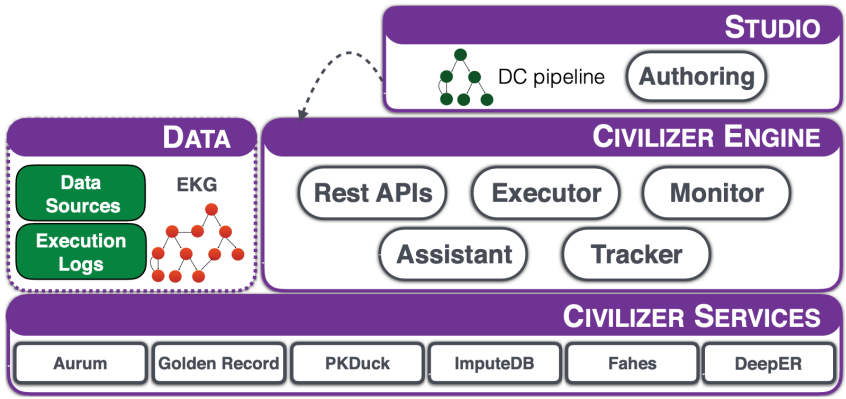}
\caption{\emph{Data Civilizer architecture}}
\label{fig:dc1_architecture}
\end{figure}

As part of our Data Civilizer effort~\cite{Deng17} we have developed modules that deal with the most common data integration tasks. Figure~\ref{fig:dc1_architecture} illustrates the architecture of \emph{Data Civilizer}. Users author their data integration workflows using the Studio. The building blocks of the workflows are pre-packaged data integration modules that were designed to target the most common pain points in data integration. The Engine then runs the workflow.

We briefly describe each module and the task it addresses.

\noindent{\textbf{Data Discovery}}: First off, data scientists need to find relevant data tables from disparate data sources (e.g., data lake) that they could use for a particular task (e.g., find customers who have been making timely payments over the past 5 years). This step is referred to as \textit{Data Discovery}. Data Civilizer includes a module, $Aurum$~\cite{aurum}, that allows users to find relevant data tables given various queries (e.g., is there a primary key - foreign key (PK-FK) relationship between the customers table and the payment history?).

\noindent{\textbf{Data Preparation}}: This step, also known as Data Cleaning, encompasses all the data transformations that are needed to bring the data to a ``usable'' format. Examples of data preparation include data standardization (e.g., use CS instead of Computer Sci. or Computer S.) and filling in missing values. We now present some of the data preparation modules we have developed in \emph{Data Civilizer}.

\begin{itemize}
\item PKDuck~\cite{pkduck} performs approximate string joins to map abbreviations with their full forms in a given set of table columns. PKDuck helps standardize values representation for downstream processing by replacing the full string forms with their corresponding abbreviations.

\item ImputeDB~\cite{imputedb} performs query-time missing values imputation. Given a query, ImputeDB generates a query plan that includes imputation operators to efficiently impute missing numerical values using any plugable statistical imputation technique.

\item Fahes~\cite{fahes} detects disguised missing values (DMVs) in the data using outliers and inliers detection techniques for categorical and numerical data. Fahes replaces DMVs with NULL in each input table.

\item DeepER~\cite{deeper} performs entity resolution using deep learning methods. Specifically, DeepER benefits from pre-trained word embeddings and user-provided training data to detect duplicate records. Additionally, DeepER makes use of blocking to reduce the number of record comparisons.

\item Aurum~\cite{aurum} helps navigating a large collection of tables to perform data discovery by building an Enterprise Knowledge Graph (EKG) that encodes relationships between those tables. Users can query the EKG to answer various data discovery questions (e.g., joinable tables).

\item A Golden Record~\cite{deng2019}  module was developed to consolidate duplicates into a single record (golden record). Before attempting to fuse the duplicate records, this module first standardizes value representations by learning candidate transformations from the data. Those transformations are then presented to the user for validation.

\end{itemize}

We now move on to the latest version of \emph{Data Civilizer}, which, in addition to the aforementioned data integration features, includes major improvements  geared towards making the jobs of data scientists easier.

\newpage
\section{Bridging the Gap between Data Preparation and Data Analytics}

Data scientists spend the bulk of their time cleaning and refining data workflows to answer various analytical questions.  Even the most simple tasks require using a collection of tools to clean, transform and then analyze the data. When a machine learning model does not produce accurate results, it is due to (1)~raw data not prepared correctly (e.g., missing values); or (2)~the model needs to be tuned (e.g. fine-tuning of the model's hyperparamters).
While there are many efforts to address those two problems independently, there is currently no system that addresses both of them holistically. Users need to be able to iterate between data preparation and fine-tuning their machine learning models in one workflow system.

We worked with scientists at the Massachusetts General Hospital (MGH), one of the largest hospitals in the US, to accelerate their workflow development process. Scientists at MGH spend most of their time building and refining data pipelines that involve extensive data preparation and model tuning. Through our interaction, we pinpointed the following hurdles that stand in the way of fast development of data science pipelines (in the sequel, we use the words ``pipeline'' and ``workflow'' interchangeably).

\sloppypar

\noindent{\textbf{Decoupling Data Cleaning and Machine Learning}}: When it comes to building complex end-to-end data science pipelines, data cleaning is often the elephant in the room. It is estimated that data scientists spend most of their time cleaning and pre-processing raw data before being able to analyze it. While there are a few emerging 
machine learning frameworks~\cite{mlflow, airflow, helix}, they fall short when it comes to data cleaning support. There is currently no interactive end-to-end framework that walks users from the data preparation step to training and running machine learning models. 
%\lei{When I read this paragraph, it is not super clear what debugging really means in this scenario.}

\noindent{\textbf{Coding Overhead}}: In larger organizations, it is typically the case that several scientists/engineers write scripts that deal with different parts of the data science pipeline. While many data science toolkits and libraries (e.g., scikit-learn) have gained a wide adoption amongst data scientists, they are only meant to build standalone components and hence are not well-suited to building and maintaining pipelines involving a wide variety of tools and datasets. As a result, scientists have to write code to build and maintain data pipelines and update the code whenever they need to refine them. Because building data pipelines is a trial-and-error process, maintaining scripts hardwired for specific pipelines is time-consuming. Moreover, the effort required to try out different pipelines typically limits the exploration space.

\noindent{\textbf{Debugging Pipelines}}: When building a pipeline involving different modules and datasets, it is typical that the final output data does not look right. This is typically due to (1)~ a problem in the modules (e.g., bug, bad parameters); or (2)~the input data to the modules was not good enough to produce reasonable results (e.g., missing values). The latter case is hard to debug using current debuggers that focus mainly on code, i.e., users have to dump and inspect intermediate data to find where it went wrong. Since it takes hundreds of iterations to converge to a pipeline that works well for the task at hand, a data-driven debugger can significantly decrease the time spent in this process. 

\noindent{\textbf{Visualization}}: Different datasets require different types of visualizations (e.g., time series, tables). Typically, scientists visualize the data in its raw format (e.g., tables) or manually visualize the data using commodity software like Microsoft Excel. However, when building pipelines iteratively, it is daunting to seamlessly integrate visualization applications (panning, zooming) into the pipeline-building process. Moreover, users need to spend a lot of time if they elect to write custom visualizations of their datasets. %We implemented 
\begin{figure}[h!]
\centering
\includegraphics[width=1\textwidth,bb=0 0 500 700]{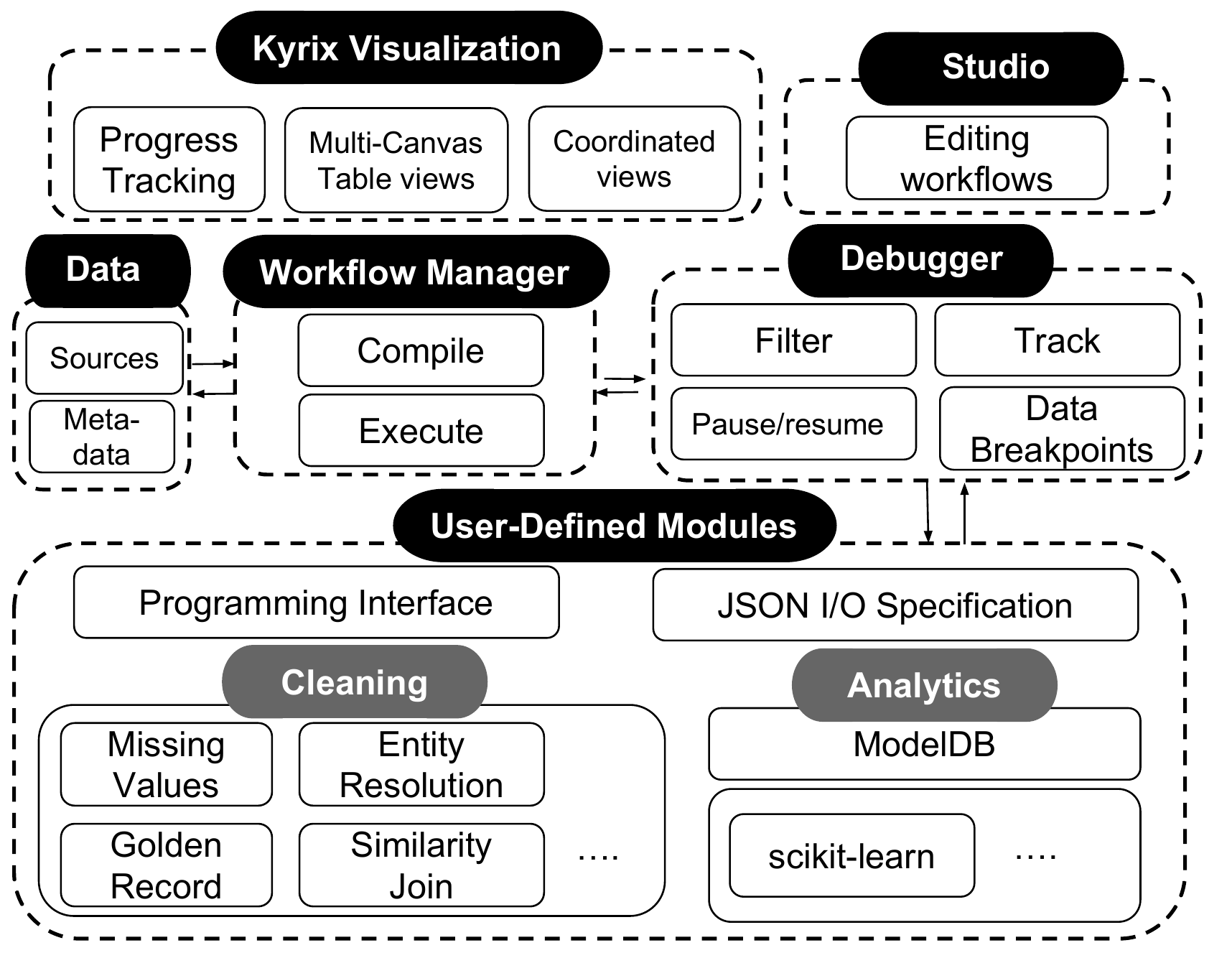}
\caption{DC2 Architecture}
\label{fig:dc_architecture}
\end{figure}

There are several efforts to support data cleaning tasks ~\cite{DBLP:journals/pvldb/PapadakisTTGPK18, magellan, holoclean}, iterative machine learning workflow development ~\cite{keystoneml, airflow, mlflow, deepdive},  and data workflow debugging~\cite{DBLP:conf/icse/GulzarIYTCMK16}. Each of those efforts focuses on one aspect of the pipeline development at a time, but not all.

The previous version of Data Civilizer~\cite{Deng17, aurum, Mansour18} focused on data discovery and cleaning using pre-defined tools. In most scenarios, users clean their data to feed it to machine learning models. 

We introduce \emph{Data Civilizer 2.0} (\emph{DC2}, for short) to fill the gap between data cleaning and machine learning workflows and to accelerate iterative pipeline building through robust visualization and debugging capabilities. In particular, \emph{DC2} allows integrating general-purpose data cleaning and machine learning models into workflows with minimal coding effort. The key features of \emph{DC2} are:

\begin{itemize}
\itemsep-1mm
\item User-defined modules: In addition to a state-of-the-art cleaning and discovery toolkit that we already provide~\cite{Mansour18}, users can also integrate 
their data cleaning and machine learning code into \emph{DC2} workflows through a simple API implementation. Users  have to simply implement a function that triggers the execution of the module they are adding.

\item Debugging: \emph{DC2} features a full-fledged debugger that assists users in debugging their pipelines at the data level and not at the code level. For instance, users can run workflows on a subset of the data, track particular records through the workflow, pause the pipeline execution to inspect output produced so far, and so on.

\item Visualization: At the core of \emph{DC2} is a component that allows users to easily implement their own visualizations to better inspect the output of the pipeline's components. We have pre-packaged a few visualizations such as progress bars for arbitrary services, coordinated table views, etc. 
\end{itemize}

\subsection{System Architecture}

We provide a high-level description of the \emph{DC2} architecture (Figure~\ref{fig:dc_architecture}) and details are discussed in the subsequent subsections. \emph{DC2} includes three core components: (1)~\textit{User-Defined Modules} cover required functionalities to support plugging-in existing user-defined modules into the workflow system (Section~\ref{execution_subsection}); (2)~\textit{Debugger} which includes a set of operations to do data-driven debugging of pipelines (Section~\ref{debug}) and; (3)~\textit{Visualization} abstractions to facilitate building scalable visualization applications to inspect the data produced at different stages of the pipeline (Section~\ref{vis}). Users interact with \emph{DC2} using the \emph{DC2} Studio, which is a front-end Web GUI interface to author and monitor pipelines.

\subsubsection{User-defined Modules}
\label{execution_subsection}
Users can plug-in any of their existing code into a \emph{DC2} workflow. Because cleaning and machine learning tools can vary widely, \emph{DC2} features a programming interface that is abstract enough to cover any data cleaning or machine learning module.  

\noindent{\textbf{Module Specification. }}
In order to specify a new module in \emph{DC2}, users must (1)~implement a \textit{module execution} function (executeService) using the \emph{DC2} Python API; (2)~load the module into \emph{DC2} by specifying its \textit{entry point file}, i.e, the file that contains the implementation of the \textit{module execution} function; and (3)~write a JSON file to list the parameters the module requires for execution.

\noindent{\textbf{Pipeline Execution. }}
Service execution happens in two phases: (1)~Studio generates a JSON object containing the authored workflow, which includes: module names, parameters and the connections between modules. This JSON object is then passed to the backend (workflow manager in Figure~\ref{fig:dc_architecture}) to run the workflow and; (2)~every module produces a JSON object containing the path of output CSV files which are then passed to the next module in the workflow. All the \emph{DC2} modules use a ``table-in, table-out'' principle, i.e., input and output of all modules is a table.  In case the module fails to run, an error code is sent back to \emph{DC2} and the pipeline execution is stopped.

\noindent{\textbf{executeService:}}
The module execution function (\textit{executeService}) takes as argument the JSON file generated from the \emph{DC2} studio. This JSON file contains the parameter values as specified from the studio for the individual modules as well as the authored workflow. Every module (1)~reads a set of CSV input files; (2)~writes a set of CSV output files; and (3)~might use metadata files if specified as an argument.

Every module can produce various output streams. We separate them into: \textit{output} and \textit{metadata}. Files produced under the \textit{output} stream are passed on to any successor modules in the pipeline while files in the \textit{metadata} stream are just meant to serve as ``logs'' that users can inspect to debug the module. For instance, a similarity-based deduplication module can produce an output stream containing the duduplicated tuples and a metadata stream that includes the similarity scores between pairs of tuples that were marked as duplicates. Each module has to produce a JSON file (output JSON) that specifies which files are produced as \textit{output} or \textit{metadata}.

\noindent{\textbf{{I/O Specification.}}}
Every \emph{DC2} module is associated with a JSON file (input JSON) containing the list of parameters the module expects and their type. Additionally, the input JSON contains the module metadata (e.g., module name, module file path). \emph{DC2} Studio needs this specification to load the module into the GUI (e.g., if a module expects two parameters, two input fields are created in the GUI for that module). 

\subsubsection{Managing Machine Learning Models}
\emph{DC2} supports adding machine learning models in the workflow. We integrated ModelDB~\cite{modeldb} into \emph{DC2} to offer first-class support for machine learning model development. ModelDB supports the widely used \textit{scikit-learn} library in Python. Users who include machine learning modules in the pipeline can (1)~track the models on defined metrics (e.g., RMSE, F1 score); (2)~implement the ModelDB API to manage models built using any machine learning environment; (3)~query models' metadata and metrics through the frontend; and (4)~track every run of the model and its associated hyperparameters and metrics.

Moreover, we have implemented a generalization of ModelDB to track metrics in any user-defined module through a light API. The \textit{DCMetric} class contains the following methods:
\begin{itemize}
\itemsep-1mm
\item \textit{DCMetric(metric\_name): } constructor which takes the name of the metric as a string (e.g., f1 score).
\item \textit{setValue(value): } sets the metric value. The metric can be set multiple times per run but only the final set value is exposed in \emph{DC2} Studio.
\item \textit{DC.register(metric): } the defined metric object is registered through this function. Registration is required so the metric is surfaced in the studio.
\end{itemize}
The following is an example code snippet to track a metric ``f1''. First, the metric is defined (line 1). Then, the metric value is set (line 2). The metric value is finally reported to \emph{DC2} (line 3).

\begin{verbatim}
1 DCMetric metric_f1 = new DCMetric("f1")
2 metric_f1.setValue(f1score)
3 DC.register_metric(metric_f1) 
\end{verbatim}

\subsubsection{Visualization}
\label{vis}
MGH datasets are massive. For instance, the one we use in this use case is ~30TB. Because we wanted to enable interactive visualizations at scale, we integrated \textit{Kyrix}~\cite{kyrix}, a state-of-the-art visualization system for massive datasets into \emph{DC2}. With Kyrix, users can write simple code to build intuitive visualization applications that support panning and zooming. The MGH scientists we worked with confirmed that visualization is a key component to make it easier for them to inspect their datasets. While users can write their own visualization applications using the Kyrix API, \emph{DC2} comes with a few generic visualization applications: (1) Progress reporting: services report their progress periodically to the Studio through a progress bar; (2) Multi-Canvas Table Views (MCTV): users can click on arcs interconnecting modules on the pipeline to visually inspect the intermediate records passing between the modules and run queries on them (e.g., filter based on predicate); and (3) Coordinated Views: in the MCTV, when users select a record in one canvas, other records are selected on other canvases based on a user-defined function (e.g., provenance, records sharing same key). \emph{DC2} comes with an API for easy integration of Kyrix visualization applications in the \emph{DC2} Studio (e.g., show a visualization application after clicking on a particular module).

\subsubsection{Debugging Suite}
\label{debug}
We have seen pipelines that run for hours, so the goal of the \emph{DC2} debugger is to catch data-related anomalies (e.g. input data is malformed in one of the modules) early in the workflow execution, so that ``bad'' data is not passed to downstream processing. \emph{DC2} features a set of human-friendly debugging operations to assist users in debugging their pipelines. We implement a GDB-like debugger that is data-driven. Users can add breakpoints by specifying a record or a set of records that satisfy predicates. Pipeline execution is paused upon reaching a breakpoint so that users can inspect visually what is going on so far in the pipeline. 
The following are the key debugging operations that \emph{DC2} provides.

\begin{itemize}
\itemsep-1mm
\item \textbf{filter}: while building a data pipeline, users typically experiment with smaller datasets before testing their pipelines on the entirety of the data. The \textbf{filter} operation allows users to specify a set of predicates to extract smaller subsets from the input datasets. For instance, if the filter is \textit{City $=$ ``Chicago''}, then, only records with City value of ``Chicago'' will be passed as input to the respective module.  

\item \textbf{track}: an important operation when refining pipelines is to be able to track a set of records to make sure the pipeline is working as expected. Users can specify filters to track records in the pipeline (e.g., track records whose \textit{City} attribute value is ``Chicago''). Whenever a record satisfies the defined filter, it is added to a \textit{tracking file} which contains (1)~the attribute values of the record before and after going through the module; and (2)~information related to the module that produced the record (e.g., name of the module, list of parameter values).

\item \textbf{breakpoints}: users can specify breakpoints in the pipeline using filters. Whenever a record satisfies the filter, the execution is paused to allow the user to inspect the record at the breakpoint. Users can then manually resume the execution.

\item \textbf{pause/resume}: this is a way for users to pause/resume the execution from the Studio. This functionality is implemented using breakpoints (more details in Sections~\ref{mb} and~\ref{ab}). This operation is useful when users only want a certain module to run for a limited period of time (e.g. pause after 5 seconds). When users inspect the output and validate it, then they can resume the execution.
\end{itemize}

\subsubsection{Manual Breakpoints}
\label{mb}
Data breakpoints serve as ``inspection'' points in the pipeline, i.e., they are used to inspect records of interest. For instance, in a deduplication module, if users notice that records whose ``City'' value is ``Chicago'' are always incorrectly deduplicated, they can add a breakpoint on records that meet the filter $City$ = ``Chicago'', then the pipeline execution is paused whenever a record that meets the filter is encountered.  We provide an API to allow users to programmatically define functions to set data-driven breakpoints. Those functions are used by the \emph{DC2} Studio to allow users to interactively set breakpoints on records that satisfy a given user-provided filter. Three key functions need to be implemented in the \textit{entry point file} (file containing the \emph{DC2} API implementation) to enable manual breakpoints: (1)~\textit{setBreakpoint} which takes as argument a filter (e.g., $City$ = $``Chicago$''); (2)~\textit{pause} to pause the execution when a record satisfying the filter is encountered; and (3)~\textit{resume} to resume the execution after the user has inspected the records on the breakpoint.

\subsubsection{Automatic Breakpoints}
\label{ab}
In some cases, implementing the API to enable manual breakpoints can be time-consuming. To address this hurdle, \emph{DC2} can create breakpoints in modules automatically (i.e., without requiring users to implement an API). This is done by partitioning the input data (of the module) into different subsets and running the module with each partition. The goal is to be able to detect errors in the output of the module run with fewer partitions than with the entirety of the data. For instance, when running a classification module (with an already trained model), users might want to inspect the output for every 10\% of the input data which results in nine breakpoints, i.e., output is shown after 10\%, then after 20\%, and so on. Additionally, the classification label of a given record does not change whether we run the model with the entire data or only a partition. If users detect misclassified records with a run using 20\% of the input data, then, there is no reason to run the module for the remaining 80\% records. Moreover, users can specify predicates to create partitions (blocking). For instance, ``City = *'' would create partitions (or blocks) where records in the same partition share the same value of the ``City'' attribute. Users can create automatic breakpoints from the \emph{DC2} studio.

\begin{figure}[h!]
\centering
\includegraphics[width=10cm,bb=0 0 700 500]{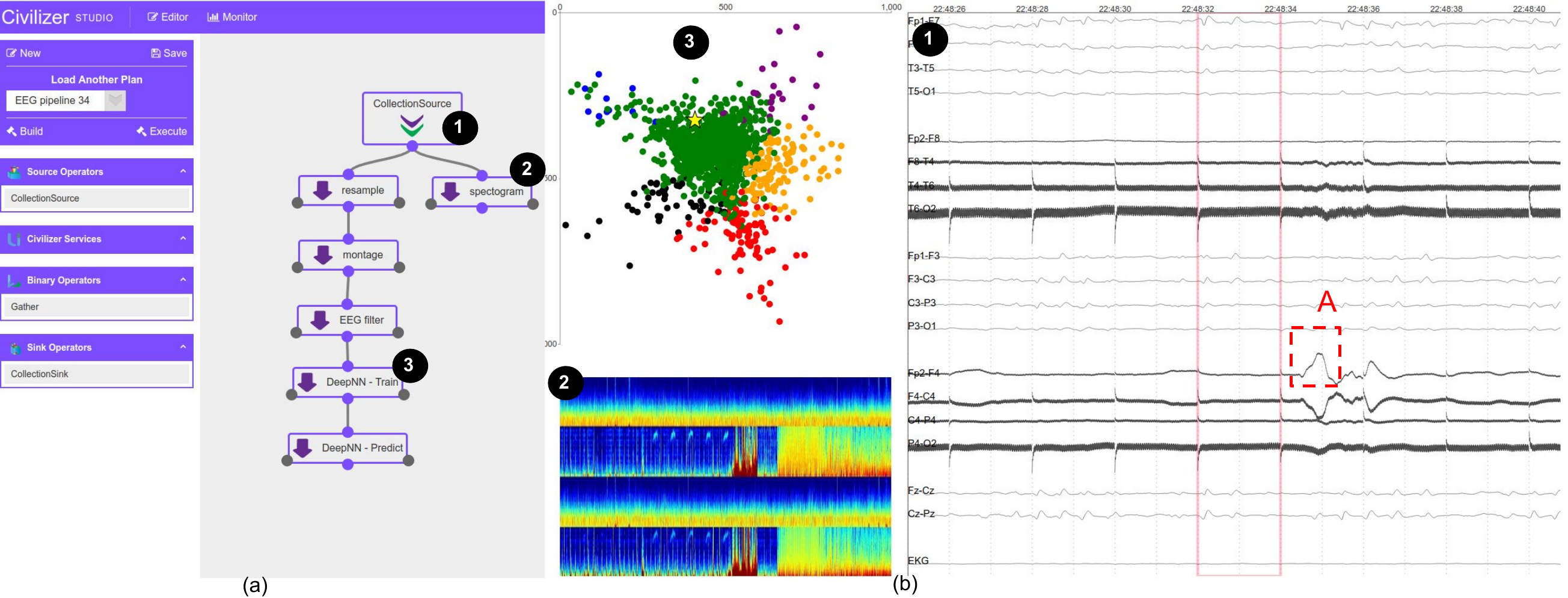}
\caption{(a)~EEG pipeline example. (b)~Visualization of numbered components.}
%\vspace{-2pt}
\label{fig:screenshot}
\end{figure}

\newpage 

\section{Data Civilizer Use Case}
We used \emph{DC2} in a real-world a medical use case with a group of scientists at MGH studying brain activity data captured using electroencephalography (EEG). Figure~\ref{fig:screenshot}(a) illustrates an example pipeline to clean the EEG data before running it through a machine learning model. In Figure~\ref{fig:screenshot}(a), each numbered module in the pipeline has its corresponding visualization in Figure~\ref{fig:screenshot}(b) (e.g., module numbered 1 corresponds to raw data input).

\noindent{\textbf{Study.}} Scientists at MGH start with a study goal (e.g., early detection of seizures using EEG data), and then prepare the relevant datasets using cleaning modules. They then apply machine learning models to perform a prediction task. In the case of this use case, they want to predict seizure likelihood given EEG labeled segments. This process is iterative in nature and  it takes several iterations to converge to a ``good'' data pipeline.
We helped the MGH scientists clean and then analyze the EEG data using machine learning models. We will walk the audience through how \emph{DC2} was used to help  quickly design and execute data pipelines to carry out the study at hand.

\noindent{\textbf{Dataset.}}
The EEG dataset 
pertains to over 2,500 patients and contains 350 million EEG segments. The total dataset size is around 30TB. Active learning is employed to iteratively acquire more and more labeled EEG segments as described in the scenario below.

\noindent{\textbf{Scenario.}}

The workflow scenario goes as follows: 
(1) Raw EEG data is cleaned. In addition to the cleaning toolkit that comes with \emph{DC2}, we plugged the cleaning tools MGH scientists use to clean the data into \emph{DC2} as user-defined modules. An example cleaning task is to remove high-frequency signals (e.g., area A in Figure~\ref{fig:screenshot}(b)); (2) Using the visualization component of \emph{DC2}, the specialists interactively explore the 30T EEG data and then label the EEG segments based on their domain knowledge; (3) After acquiring a set of manually labeled segments, a label propagation algorithm, as a user-defined component of \emph{DC2}, automatically propagates labels to the nearby segments of the existing labeled segments; (4) A deep learning model is then learned using part of the labeled segments as training set. During this process, the \emph{DC2} debugger is fully explored to tune the hyper-parameters and the network structures; 
(5) Active learning is then conducted to improve the quality of the automatically acquired labels. First, the labeled segments out of the training set are classified by the learned model. Then using the ModelDB component of \emph{DC2} the 2000 segments are efficiently extracted where the neural net had highest confidence but disagreed with the labels; (6) These segments are then fed back into the visualization component for the domain experts to decide whether they need to update their labels (go back to step 3) or review the cleaning step (go back to step 1). This iterative process proceeds until the neural net reaches a satisfactory classification accuracy.

\newpage

\section{Knowledge Graphs}

\subsection{Background and Terminology}
\label{sec:backgroundkn}

Knowledge Graphs --- sometimes called a Knowledge Network or 
structured Knowledge Bases --- are a relatively novel way to describe
structured data about a particular topic.  Examples include Wikidata (\cite{wikidata}),
DBpedia (\cite{dbpedia}), the Google Knowledge
Graph (\cite{singhal_2012}), UniProt (\cite{uniprot}),
MusicBrainz (\cite{musicbrainz}), GeoNames (\cite{geonames}), and many
others (\cite{suchanek:2007, knowitall, bizer:2009}).  They have had slightly
different definitions over the years.  For the sake of this document,
we think of a Knowledge Graph as a data resource in which:
\begin{itemize}
  \item There exists a set of {\em unique entities} that correspond to
        real-world objects.  For example, entity Q76 represents Barack Obama
        in the Wikidata knowledge network, and entity Q6279 represents
        Joe Biden.  Different knowledge graphs make different
        decisions about what entities should be contained.  For
        example, there is a Joe Biden entity in Wikidata, but not in the
        MusicBrainz knowledge network, which specializes in recorded
        music\footnote{Because Barack Obama recorded the audiobooks for his
        self-authored books, he appears in both Wikidata and
        MusicBrainz.  His MB identifier is {\tt
          0de4d19f-05c8-4562-a3c0-7abdc144f1d5}.}.  These can be
          thought of as the nodes in the knowledge graph.

  \item There are {\em unique properties} that describe a relationship
        between an entity and a data value.  For example, Wikidata
        property P19 describes the {\em place of birth} relationship.
        For most knowledge graphs, properties are expressed with two
        values, at least one of which must be an entity.  For example,
        the {\em place of birth} property describes a
        relationship between two entities in the knowledge graph,
        one of which is usually a person, and the other usually a
        location.  In contrast, Wikidata's property P569 ({\em
        date of birth}) usually describes a relationship between a
        single entity and a data value.  These can be thought of as
        potential edge labels in the knowledge graph.

  \item By combining entities, properties, and data values, the
        knowledge graph asserts a large number of true facts about
        real-world objects.  For
        example, Wikidata states that (Q76, P26, Q13133) is true.
        That is, Barack Obama (Q76) {\em has spouse} (P26) of
        Michelle Obama (Q13133).  These triples can be thought of as
        concrete labeled edges in the knowledge graph.
\end{itemize}

Our definition for a knowledge graph places it outside the
ambiguity and reference problems often caused by natural language.
For example, Wikidata properties P19 ({\em place of birth}) and P569 ({\em date of
  birth}) could both be described in English by the phrase ``born in,'' but
  this point is irrelevant for the knowledge graph's correctness.
  There are two real-world and distinct relationships, so there are
  two distinct properties in Wikidata.  Some applications, such as
  voice assistant question answering, might have to manage linguistic
  ambiguity, but that challenge is the responsibility of the
  application, not the knowledge graph.

There is nothing preventing us from using KGs to contain entire
datasets rather than single scalar values.  For example, it could make sense to create an entity to model a
well-known GDP dataset for practicing economists, which has multiple
columns and rows; or a set of vessels when managing a fleet.  Indeed, Wikipedia already does something similar
when its entities contain datasets, as in the case of the Wikipedia
entity {\tt Economy\_of\_the\_United\_States}, which contains annual
stats for GDP, inflation, unemployment, and so on.

\begin{figure}[h!]
  \centering
  \includegraphics[width=1\textwidth,bb=0 0 1300 500]{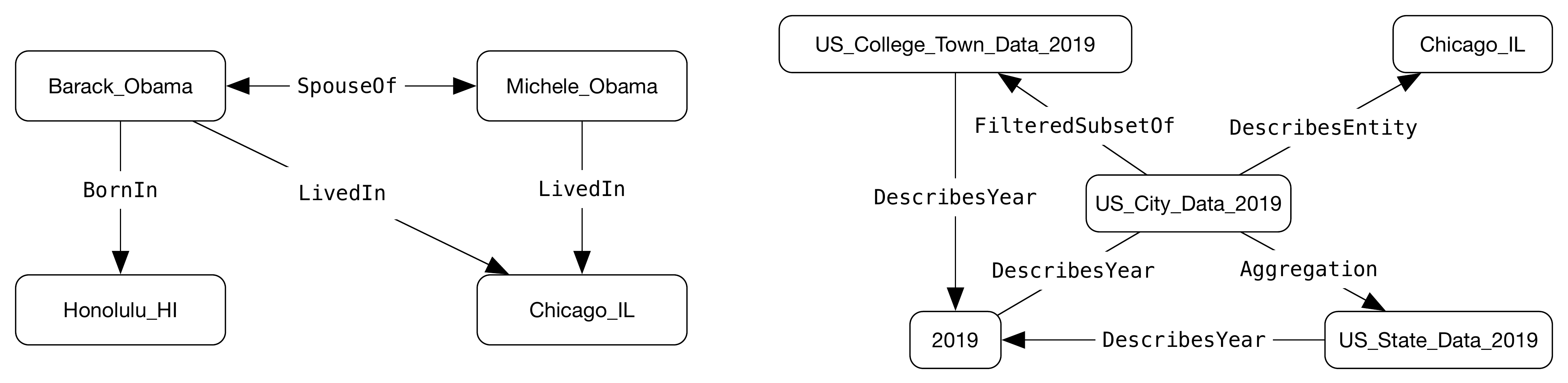}
  \vspace{-0.2cm}  
  \caption{On the left, a fraction of a general-purpose knowledge network.  On the
    right, a fraction of an Economics-specific knowledge network.}
  \label{fig:kg}
  \end{figure}

Figure~\ref{fig:kg} shows two small example knowledge graphs. The figure on the left shows a portion of a general-purpose knowledge graph. The edge labels appear in many edges in the graph. The figure on the right shows an Economics-specific knowledge graph.

Our definition is not entirely consistent with all of the 
academic literature.  There are academic knowledge graphs that
could potentially create a node for every distinct noun phrase in a
text, even if they refer to identical real-world objects, such as
VerbKB (\cite{verbkb:2016}).  However, our definition is consistent
with the major deployed knowledge graphs, such as Wikidata, DBpedia, and many
others\footnote{Linguistically-driven
networks like VerbKB are useful in some cases.  For example,
Biperpedia is a ``best effort'' data resource extracted from text that
can help search applications (\cite{gupta:2014}).  Also, there is a
growing literature in the NLP area that attempts to use text directly
for question answering (\cite{squadOrig, squad, bidaf}), without producing a knowledge network at all,
but these efforts today are primarily still in the research
sphere and it is unclear how to extend them to non-question-answering applications.}.

\subsection{Graph Databases}
It is useful to distinguish between a {\em knowledge graph} and  {\em graph database software}. Graph database systems, such as Neo4j~\cite{webber2012programmatic} or Amazon Neptune~\cite{bebee2018amazon} offer query and update services for graph-oriented datasets, much like relational database systems do for traditional relational datasets. Just as traditional database software like Oracle offers the SQL relational query language, graph databases will offer a query language tailored for graphs. The most popular language is probably SPARQL, although Neo4j's Cypher language is both well-known and especially relevant when processing knowledge graphs.

Like SQL, graph query languages enable the user to ask precise and interesting analytical questions, but most users do not have the training needed to write them. For example, here is a simple query (drawn from the World Wide Web organization's SPARQL tutorial) that obtains a list of 50 spacecraft from a knowledge graph:

\begin{verbatim}
PREFIX space: <http://purl.org/net/schemas/space/>
SELECT ?craft
{
  ?craft a space:Spacecraft
}
LIMIT 50
\end{verbatim}

It is possible to understand this code, but writing even this simple query will require user training.

Graph databases are designed to store a wide variety of graph-structured information, not just knowledge graph data. They are also useful for storing social network data, hyperlink data, bioinformatics data, and other kinds. Indeed, most applications of graph databases do not involve knowledge graphs. Knowledge graphs are usually of fairly modest size compared to other kinds of graph-structured data. Consider that as of this writing, Facebook has about 2.5 billion users in its social graph, while Wikidata has only 81 million entities in its knowledge graph.

Graph database systems and knowledge graphs are sometimes confused, but they play different roles. A knowledge graph is a certain kind of dataset, while a graph database system is a piece of software. Although they are often useful together, one cannot be substituted for the other.

\subsection{Application Examples}

There has been a substantial amount of research into methods for construct knowledge graphs, such as information extraction~\cite{gatterbauer2007,suchanek:2007,snorkel, deepdive, cafarella:2008b, knowitall, resolver, webtables:2015} and crowdsourcing ~\cite{Kobren2014DomainSK,Kobren2017EntitycentricAF, angeli-etal-2014-combining, angli,angeli-etal-2014-combining}.

In contrast, there has been relatively little research into the knowledge applications built on top of KGs, outside a few important but narrow use cases, such as personal assistants \cite{fader:2014} and precision medicine \cite{precisionmedicine}.

\begin{figure}
    \centering
    \includegraphics[width=0.8\columnwidth,bb=0 0 900 700]{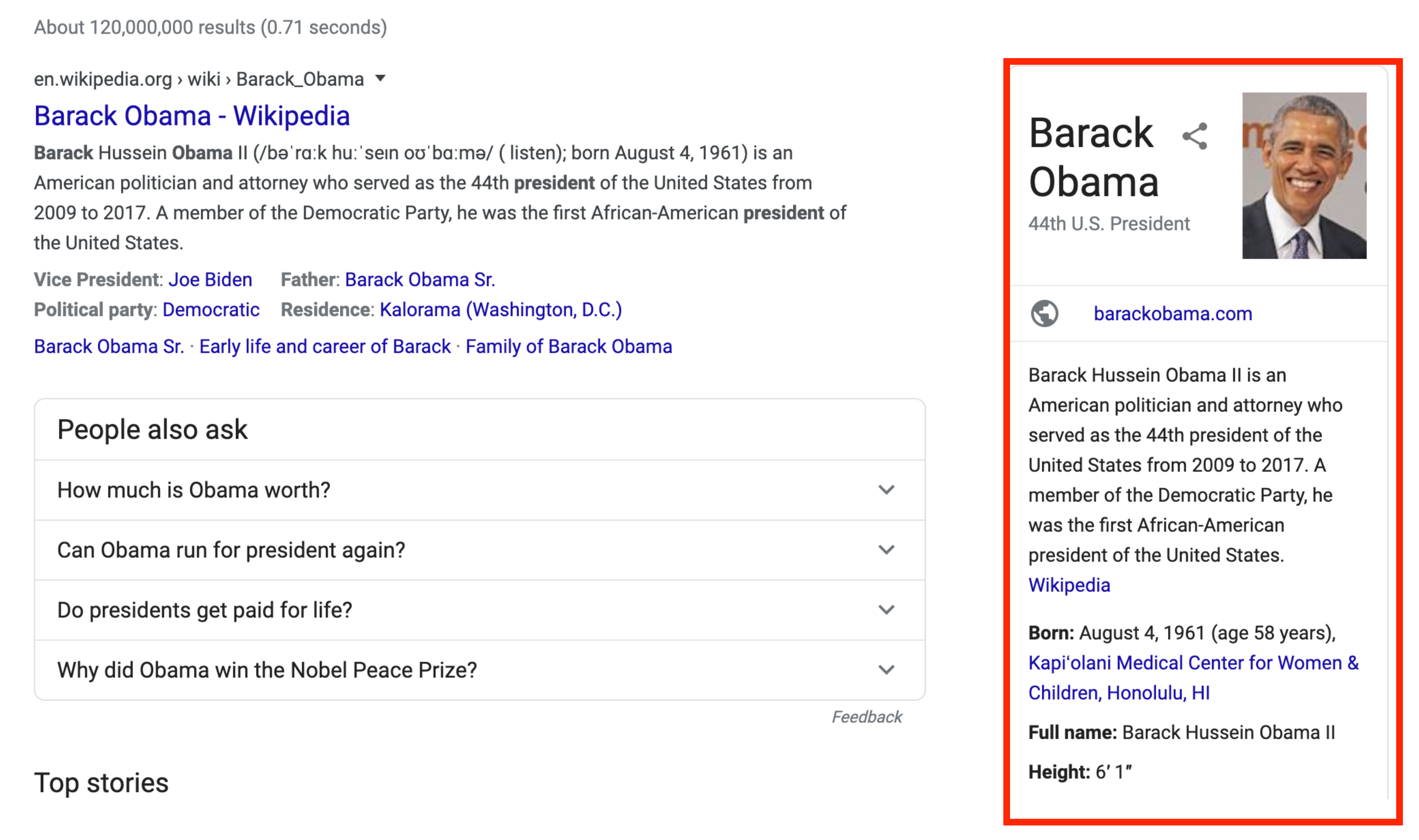}
    \caption{A Google search yields the traditional list of hyperlinks at the left, and often a knowledge graph-derived structured results on the right-hand side. In this case, {\bf born}, {\bf Full name}, and {\bf Height} are all properties of the {\bf Barack Obama} entity in the KG.}
    \label{fig:structuredwebsearch}
\end{figure}

The best-known industrial use cases of knowledge graphs are {\em structured web search} (see Figure~\ref{fig:structuredwebsearch} and {\em voice assistants} such as Siri, the Google Assistant, and Amazon's Alexa product. In both types of systems, a knowledge graph serves as a generic multitopic database to produce user answers.

It is perhaps useful to consider the rough pipeline of steps that take place in a voice assistant to understand the role that a knowledge graph plays. Imagine that a user has just asked the Google Assistant, "Where was Barack Obama born?" Conventional engineering wisdom suggests the voice assistant goes through the following steps:
\begin{enumerate}
    \item The device records user speech saying, "Where was Barack Obama born?"
    \item Voice assistant software uses a speech-to-text system to translate the utterance into a textual string
    \item The voice assistant applies natural language processing methods to the text string in order to translate it into a structured graph query in SPARQL or something similar.
    \item A graph database system that stores the knowledge graph will execute the SPARQL query and return a result.
    \item The voice assistant will turn the result into audio that can be played to the user.
\end{enumerate}

A large and high-quality knowledge graph is necessary, but not sufficient, for a successful voice assistant. The voice assistant can only answer questions that have answers in the knowledge graph, which potentially explains Google's private efforts to create a very large one~\cite{singhal_2012}. 

\subsection{Wikidata}

Wikidata is a widely-used open-source knowledge graph. Wikidata was founded in 2012 and has since become the largest and most successful of the public general-interest knowledge graphs. Wikidata is unusual in being both very large and very transparent. In contrast, most other knowledge graphs are either small research projects or large private endeavors. It is a volunteer nonprofit effort, with funding donated from various charitable and scientific organizations. 

As of this writing, Wikidata contains structured data about roughly 82M entities. An entity is any general-interest real-world object, such as Barack Obama (\url{https://www.wikidata.org/wiki/Q76}), the city of Chicago (\url{https://www.wikidata.org/wiki/Q1297}), the Federal Reserve (\url{https://www.wikidata.org/wiki/Q53536}), or the US Navy (\url{https://www.wikidata.org/wiki/Q11220}). Wikidata contains approximately 1 billion factual statements. Roughly half of the entities (40M) contain 10 or more factual statements.

Wikidata is remarkable in its ability to simultaneously obtain large scale and data quality. Not only is the factual precision high, the set of properties and types --- Wikidata's version of a schema --- is remarkably consistent across the dataset. For example, there is a single property {\em spouse} that is widely used; there are few or no unnecessary duplicates of the concept. This might not sound like a major achievement, but consider much overlap and duplication exists in all the relational database schemas in an organization. Even a simple concept like {\em employee} can be modeled in many different ways in different databases; if the organization wants to run a query that examines {\em all} of the employees, it often must perform a long and expensive data integration process. Wikidata is somehow able to avoid most of these modeling and integration problems, even though its dataset and topic diversity are vast.

It is not entirely clear which Wikidata practices are most responsible for their success, but here are a few notable ones:
\begin{enumerate}
    \item Almost any user can contribute a new fact to Wikidata, without any scrutiny before it is added and made publicly visible. However, they are limited to using properties (e.g., {\em spouse} or {\em sibling} or {\em date of birth}) that already exist.
    \item Any user who wants to add a novel property must submit the request for approval before the property can be used in a new fact. The request is reviewed by a small panel of senior Wikidata editors.
    \item The software interface for adding a novel fact includes aggressive autosuggest, which encourages users to select properties from the preexisting set.
    \item The Wikidata software retains full version history of all facts, so it is easy to undo any unwanted changes.
    \item Editors have tools that allow them to quickly review large numbers of factual additions, letting them examine and potentially undo fact insertions with just 1-2 keystrokes.
    \item The system, somewhat surprisingly, does not use much in the way of machine learning-style probabilistic machinery.
\end{enumerate}

Unfortunately, there is no commercial product that allows a user to easily replicate the Wikidata process when building a novel knowledge graph. However, it is clearly an exciting and relevant example that can potentially be emulated.

% \subsection{Related Work}
% {\em what the field has been working on recently}

\subsection{Future Research}
By massively expanding the availability of well-administered structured data, and by allowing anyone (not just database administrators) to improve that structured data, knowledge graphs promise a revolution in the kinds of data-powered applications that can be constructed. Everyday users could stop issuing simple document ranking queries; instead, with the same trivial amount of work, they could easily ask detailed questions about the world and receive a customized data-driven answer. Professional data analysts could easily focus on novel topics, not just the databases that have been cleaned and prepped for them. Visualizations in a report or a news article could be tied to their backing data stores, allowing any reader to explore a data point that seems suspicious. 

%{\bf MIKE: Vijay, maybe a Navy-specific example here? What do you think would be compellig?}

More concretely, consider:
\begin{itemize}
    \item An knowledge graph that contains virus-related scientific research, when combined with an KG-powered application, could allow a user to quickly identify the researchers who have authored the most papers that mention both remdesivir and a filovirus. 
    \item An knowledge graph that describes naval vessels, their crew ({\bf terminology?}), and their equipment could power an application that finds repair opportunities.
    
    %\item {\bf MIKE: A few more Navy apps would be great }
\end{itemize}

This exciting future has arrived in small glimpses: voice assistants, and search engines that give direct answers instead of forcing users to trawl through documents. Unfortunately, knowledge graphs do not exist for many topics, and KG-powered applications exist for fewer. The exciting knowledge graph future has been frustratingly slow to arrive. We need domain-independent infrastructure to hurry it along.

\vspace{0.2cm}
\noindent {\bf Knowledge Graph Construction} ---  One core reason is that constructing a high-quality KG itself is difficult and time-consuming.  Google has been developing its private knowledge graph for over a decade; and even though the Wikidata project is one of the best KGs available, it is almost eight years old but still has quality and coverage challenges. These are large projects, and so perhaps have necessarily taken a long time. But in past interviews with KG researchers, we found that even a small new knowledge graph project typically takes 12 months, even with highly-trained Ph.D.-level engineers. 

\vspace{0.2cm}
\noindent {\bf Knowledge Applications} --- The pain associated with building applications for knowledge graphs is even more acute. Only a tiny number of the most resource-rich organizations have been able to pursue them. Consider that building an KG-powered application entails: 
\begin{itemize}
    \item Writing complicated data acquisition code that clumsily exports data from a KG, then reimports it. Worse, this code simply breaks when the KG changes, even when it {\em improves}. 
    \item Fixing bugs that can hide in many more places than in traditional software:  bugs can lie in source code in the application itself, or incorrect inputs, or incorrect shipped data from collaborators.
    \item Shipping massive datasets to collaborators, and finding data quality error induced by downstream software.
\end{itemize}    

Moreover, because KGs and their applications have a symbiotic relationship --- better applications raise the demand for good KGs, and better KGs enable better applications --- the huge cost of developing applications means that KGs develop more slowly than they otherwise might. Unsurprisingly, the recent explosion of tech company investment in improving general-purpose KGs has gone hand-in-hand with increased popularity of applications like the Google Assistant.

\vspace{0.2cm}
\noindent {\bf Research Goals} --- We propose to research and build infrastructure that will make KGs and applications dramatically cheaper and easier to build. We plan to do so in two ways:
\begin{itemize}
    \item Building a Knowledge Graph Programming System, or KGPS. This is a programming language and toolset specifically designed for building KG-driven software and for debugging data quality problems more quickly and at lower cost than is common today.
    \item Building and rapidly shipping KGs for several applications, in order to gain experience and test integration with the KGPS.  We will initially target COVID-19 science, plus related applications. Better scientific understanding of the virus is an urgent social need that this KG and its applications can help address, while also serving as a testbed for our infrastructure. 
\end{itemize}

We can now describe our plans for the KGPS in more detail.

\subsection{The Knowledge Graph Programming System}
\label{sec:knps}

The Knowledge Graph Programming System is a programming system that takes the KG as its core data representation. It makes KG applications easier to build and easier to improve. Its features fall into three main areas:
\begin{enumerate}
    \item {\bf Easier Knowledge Acquisition} --- \system\ application code can use values and types that cover the "real world" just like a KG does. There is a value for {\tt Barack Obama}, one for {\tt Boston Red Sox}, and so on. There is a type for {\tt Politician}, for {\tt Cartoons}, {\em etc}. Unlike most programming platforms, the \knps\ library is intended to have the same comprehensive coverage that a good KG does. As a result, programmers can very succinctly obtain data from a backing KG.
    \item {\bf Easier Knowledge Debugging} --- Every output value that \system\ has a globally-accessible URL that carries the full history of how the value was computed. Many applications today are a mixture of data and code collected from a hodgepodge of colleagues, models, and opaque institutions; as a result, debugging an incorrect output can turn into a tedious goose chase rather than an engineering exercise. In contrast, \knpl\ application values will automatically have full lineage information that makes them debuggable by every programmer.

    \item {\bf Easier Knowledge Sharing} --- Data pipelines are everywhere in modern data science, including model training procedures, crowdsourcing updates, web crawlers, sensor readings, market updates, data cleaning methods, or just bureaucratic processes. Each of these pipeline steps can entail heavy data transfer costs. Worse, errors can be induced by pipelines entirely outside the application's control. \knps\ programs will have data sharing operations as a builtin primitive, making it easier to collectively diagnose and repair knowledge applications, even if those errors are tied to errors from upstream or downstream pipelines. 
\end{enumerate}

Finally, the \system\ system also has one very boring quality: it is mainly Python, so data scientists and programmers can use it with little change to their daily routines.

\subsection{Example Walkthrough}
\begin{figure}
    \centering
    \includegraphics[width=\linewidth,bb=0 0 200 100]{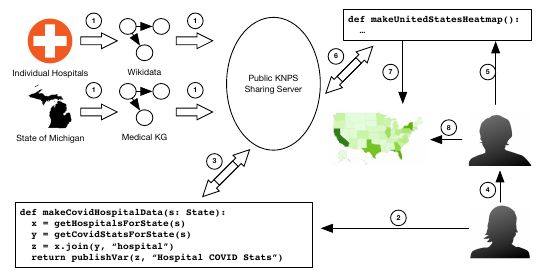}
    \caption{An example of how the \fullknps\ enables efficient and high-quality knowledge applications in the case of COVID-19 analysis.}
    \label{fig:narrative}
\end{figure}

It is perhaps easiest to describe \system's behavior by showing a simple walkthrough example. The following narrative discusses steps illustrated in Figure~\ref{fig:narrative}:
\begin{enumerate}
    \item In {\bf Step 1}, various hospitals over time add statistics about themselves (date of founding, number of beds, etc) to the public Wikidata Knowledge Network.  The State of Michigan regularly publishes the latest per-zipcode COVID case numbers data to a public health KG.  All updates to these KGs are monitored by a {\bf \system\ sharing server} and turned into variables and values that are available to all \system\ programmers.
    \item In {\bf Step 2}, a programmer writes a short \system\ function that computes the number of COVID cases per hospital bed in each zipcode in a state, applying it to the value {\tt Michigan}. Because all of the data is available via the \knps\ server, the programmer can load it with a single unambiguous function call, rather than slinging CSV files or SQL for private databases.
    \item In {\bf Step 3}, the program runs and computes the dataset of cases-per-bed. The result is both returned to the programmer and is synchronized with the \system\ server. The \system\ server returns a URL that can be used to permanently refer to the generated dataset.
    \item In {\bf Step 4}, the programmer sends this URL to a medical data analyst at a different institution, so she can experiment with this new dataset.
    \item  In {\bf Step 5}, the medical data analyst writes a short \system\ function that uses this Michigan cases-per-bed dataset to build a national heatmap of hospital usage intensity. This program refers to the dataset using the URL sent by the programmer.
    \item In {\bf Step 6}, the heatmap program runs.  It needs the programmer's data in order to run, so it contacts the \system\ server and downloads the data.
    \item  In {\bf Step 7}, the heatmap program completes and generates its output image. The medical data analyst examines the image and finds a surprising result in the heatmap near Ann Arbor.
    \item In {\bf Step 8}, the medical data analyst examines the entire lineage of how the heatmap was computed, including the original programmer's code, and any crowd modifications to the source KGs. She realizes that the number of hospital beds in Ann Arbor is incorrect, so she edits Wikidata to fix the problem. Because she has access to the full history of how the hospital data was computed, it is trivial to reexecute the entire chain of programs and to obtain a satisfactory heatmap.
\end{enumerate}

This short example illustrates many ways \system\ can make KGs more useful and knowledge workers more productive. The programmer in Step 1 can obtain an accurate list of US hospitals in just one line of code. The medical data analyst can easily examine how her programmer friend computed the dataset she received in Step 4. She can quickly find the source of the bug she identified in Step 7's output. The medical data analyst in Step 8 can directly fix the upstream data bug in Wikidata, so that everyone can benefit from the fix. 

\subsection{Language Features}
We now go through each of the key \knps\ features.

\subsubsection{Types and Values}
A core design goal of the \knps\ is to make it easy for programmers to acquire, use, and update KG data. As a result, KG entities are a primitive data value, and KG-derived types are built in to the \knps. In other words, programs are not limited to describing the world in terms of integers, strings, and sorted arrays. Rather, programmers can write code that models the real world much more directly.

As a result, programmers can easily identify sets of data to load without handling files, formats, or SQL. They can also exploit a high-quality preexisting set of schema decisions for real-world objects; these are derived from a source KG. With the \knps\ value and type system, much of the data loading and management code that is done by data software can effectively be offloaded onto preexisting KG processes.

\begin{figure}
    \centering
    \begin{verbatim}
1        def computeStandardBedsPerResident(h: Hospital):
2          nearbyZipcodes = Zipcode.getAll().filter(zc: zc.distance(h.zipcode) < 40)
3          nearbyPopulation = 0
4          for z in nearbyZipcodes:
5            nearbyPopulation += z.population
6          return nearbyPopulation / (h.beds - h.icuBeds)
7
8        computeStandardBedsPerResident(KNPS.get("Univ of Michigan Hospital"))
    \end{verbatim}
    \vspace{-0.3cm}
    \caption{KGPS code to compute the number of residents per bed in a hospital's service area.}
    \label{fig:codeexample}
\end{figure}

Figure~\ref{fig:codeexample} illustrates \knps\ code that computes the number of standard hospital beds per resident offered by a particular hospital for a standard service area. On line 1, the {\tt Hospital} type is part of the \knps\ builtin class library, derived from a backing KG.  This {\tt Hospital} type has various useful fields that have been developed and populated by the KG's data curation process, so the programmer does not have to. On line 2, the programmer uses the builtin {\tt Zipcode} type to get a list of all zipcodes, then filters them to retain only the zipcodes within 40 miles of the hospital's zipcode. The {\tt Zipcode.distance()} function and the {\tt Hospital.zipcode} field are all part of the standard \knps\ library. Many other useful fields are also part of the KG-derived types: the {\tt population} field on line 5, and {\tt beds} and {\tt hospitalBeds} on line 6. Finally, on line 8, the programmer grabs a data value that represents the University of Michigan Hospital, which is an instance of the {\tt Hospital} type.

\vspace{0.2cm}
\noindent {\bf Broader Impact} --- Without the \knps\ builtin values and types, the code in Figure~\ref{fig:codeexample} would be substantially harder to write. Programmers would have to build a custom {\tt Hospital} class and define its fields, build or download a {\tt Zipcode} class and define its {\tt distance()} function, then find data for the University of Michigan and hospital and load the data into the appropriate fields. \knps\ relies on KG-derived data to do all of that tedious, but crucial, data modeling work.  

\vspace{0.2cm}
\noindent {\bf Intellectual Novelty} --- 
The \knps\ builtin library is unlike most programming language libraries: it is intended to cover {\em every topic}, so the programmer rarely has to worry whether a particular concept is part of the system; she can simply assume it exists. We pursue this by asking the developer for a tiny number of sample instances for each desired class, then exploit recent advances in knowledge graph embeddings (such as \cite{xie:2016}, \cite{wang:2019}) to automatically build a type-specific prediction model. However, the type library cannot shed previously-defined fields or functions without potentially breaking \knps\ user code; this challenge is quite similar to problems in data integration problems (\cite{beaver}, \cite{cafarellamultiresolution}, \cite{samplemapping}). 

\subsubsection{Automatic Provenance}
Another core design goal of the \knps\ is to make all data values easy to debug. Even today's limited knowledge applications entail a vast number of moving parts and the efforts of many different people. For example, answering a single voice assistant query might depend on a fact added to the knowledge network by a Wikidata user, a piece of voice recognition code added by an engineer on an open source recognition project, a large set of training examples contributed by users, and interface code written by the company that made the voice device. If you get a wrong answer to your voice assistant query, which of the above elements was to blame?

This crucial data debugging is in some ways similar to traditional software debugging, except that the full program execution is spread out in time, across many machines, and performed by many different participants. Without knowledge of the full history of any particular piece of data, fixing errors is not possible. Developers must perform tedious detective work and track down every possible input, or in many cases simply add more training data to the model process and hope for the best.  Although there has been some recent research into systems for data debugging, it has focused on debugging a database in isolation, not data in the context of a multicomponent application (\cite{dagger}).

\knps\ addresses this challenge by giving {\em every single value} the full provenance of how it was computed. There has been a substantial amount of research in data provenance --- sometimes called data lineage --- in a database or reproducibility setting (\cite{survey_on_provenance_17, AVOCADO, vistrails, 10.1007/978-3-540-85259-9_6, 10.5555/2889875.2889881, 10.1145/1247480.1247631, 10.1145/2882903.2915248}). But there is no standard programming system in which provenance plays a major role. \knps\ aims to perform this service. 

There is a huge number of existing provenance systems, but they tend to either require a substantial amount of developer adaptation, or are incomplete and leave some portion of the value history uncaptured. We will instead follow a policy of {\em automatic provenance capture} and {\em degrading provenance accuracy rather than coverage}. We have previously used synthesized code for data-intensive tasks to obtain higher performance in opaque user code (\cite{jahani}), or to perform data manipulation operations (\cite{foofah1, foofah2}); we will follow the same strategy with provenance capture. When dealing with software components that cannot be instrumented or wrapped, we will use machine learning models to predict the likely provenance values; it might seem irresponsible to employ potentially-inaccurate provenance values, but in a debugging context they will often be quite useable.

\vspace{0.2cm}
\noindent {\bf Broader Impact} --- \knps\ adds two features not typically seen in provenance systems. Crucially, it will compute {\em provenance without programmer interaction} and {\em across programs and storage systems}.  We aim to do this via synthesized wrappers around user-written \knps\ code. 
If successful, this approach would make provenance a practical reality for \knps\ application programmers, bringing the full history of every data value to the programmer's fingertips; as a result, data debugging would be vastly easier.

\vspace{0.2cm}
\noindent {\bf Intellectual Merit} --- There is a huge amount of work in data provenance, but also a lack of practical systems in common practice outside a few narrow use cases. This suggests {\em deployable provenance} is a promising and challenging direction.  Synthesizing wrappers is a known approach, although often done in a crude manner that simply logs the method call. We are unaware of existing work that pursues this degraded-accuracy approach for maximizing provenance coverage.

\subsubsection{Sharing Values and Variables}

Finally, \knps\ will allow easy sharing of values and variables. Running a \knps\ program will essentially generate its own synthetic "execution knowledge network", designed to cover the topic of its own execution history. Values in this network can be shared with others and commented-upon, just like values from Wikidata that are shared with the \knps\ server in Figure~\ref{fig:narrative}. This will allow \knps\ programs to easily add new items back to the shared knowledge network, which can then be easily reused by other developers.

\subsection{Alternate Interaction Modes}
\label{sec:usermodes}

The above text describes the standard programmatic interface to \knps. But once users add code and data to the system, other interaction modes are possible. In Phase 1 we built a search engine-like interface for analysts to use, which exposes \knps\ functions and data via a website. Users are able to obtain useful and interesting programmatic results with just a single line of text.

\newpage

\section{Practical Steps}

This section outlines a number of practical steps that can simplify data integration. We organize these practical steps into: 
\begin{enumerate}
    \item Data Organization
    \item Data Quality and Discovery
    \item Data Privacy
    \item Infrastructure, Technology and Policy
\end{enumerate}

\subsection{Data Organization}

\begin{itemize}

    \item \textbf{{Schema Integration:}}  Schema integration is one of the core challenges for any organization looking to query disparate datasets. A simple example: one database of employees might contain {\tt startdate} as a calendar date, while a second one contains {\tt beganservice} as a year. Writing a query that uses both of these databases entails figuring out how to map {\tt startdate} to {\tt beganservice}. In practice, solving this seemingly-simple problem can become quite daunting and expensive.
    
    Most traditional schema integration tools are intended to integrate preexisting schemas with extremely high accuracy. This makes the tools suitable for important integrated queries, such as computing payroll for people in a large organization with many employee databases. These tools often come with bundled professional services. IBM and Palantir are two well-known vendors. These projects tend to be time-consuming and relatively expensive. This is usually not a viable approach for projects that aim to make all data in the organization queryable: the per-schema integration cost is simply massive when the number of schemas is large.
    
    There are research and startup efforts that use probabilistic methods to perform schema-specific integration at more reasonable cost. Tamr is one small but growing vendor. These can be successful, but may have unusual requirements, such as extensive employee-annotated training data.
    
    More recently, knowledge graphs have employed web front-end tools and social processes to produce a single large-scale integrated schema. Essentially the knowledge graph's schema is the "target"; all raw data in the world must be formatted to fit this target or it cannot be added to the knowledge graph. This is a major achievement, especially considering the relatively low financial cost. However, it has come at the cost of some flexibility. Users who cannot accept the knowledge graph's target schema (perhaps to retain backward compatibility with legacy software) simply cannot add their data to the knowledge graph. Making matters worse, there is currently no standard commercial offering to enable private knowledge graph construction.
    
    It may sound like promulgating standardized schemas would enable the best of all worlds, but most organizations find these difficult to implement. One way to view the success of knowledge graph projects is that they are effectively social software systems that encourage organizational agreement on a standardized schema. Although there is no commercial product available today in this area, it is worth watching for in the future. Figure ~\ref{fig:schema_integration} give a high-level view of these various tools.
    
\begin{figure}
\centering
\includegraphics[width=1\textwidth,bb=0 0 500 300]{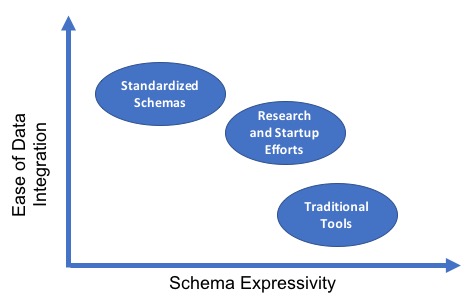}
\caption{\emph{High-level view of various tools for schema integration}}
\label{fig:schema_integration}
\end{figure}
    
     \item \textbf{{Data Formatting and Naming:}} One of the easiest ways to simplify data integration is through standardized files in human-readable formats with standardized filenames. We have observed that in many instances, a data management system can be nothing more than a shared filesystem with standardized filenames and folders. Wherever possible, parse as much data as is practical into tabular files.   Try to avoid proprietary formats.  If possible, use .csv (comma separated values) or even better .tsv (tab separated values) file formats.  Include column labels (each column label should be unique within the file).  Have row labels in the first column, such as row number or record number (each row label should be unique within the file). 
    
    Avoid lots of tiny files and compress with zip when possible. Use hierarchical directories to keep the number of items in a directory reasonable (less than 1000 files/directory).  Use well-defined directory structures to provide easy to access information about the dataset. For example:
    
    \begin{center}
    \texttt{source/YYYY/MM/DD/hh/mm/source-YYYY-MM-DD-hh-mm-ss.tsv} \\
            or \\
    \texttt{YYYY/MM/DD/hh/mm/source/YYYY-MM-DD-hh-mm-ss-source.tsv}
    \end{center}
    
  \end{itemize}
  
  \subsection{Data Quality and Discovery}

  \begin{itemize}
 \item  \textbf{{Data integrity:}} In most cases, data becomes dirty because organizations do not invest in setting up policies that control the quality of the data \textit{before} it is introduced to the database. It is always a good idea to think about constraints that should hold on the data at all times (e.g., employee salary must be greater than 0). This way, if the data is updated, we will make sure those updates would not violate the constraints.

Fortunately, there is a plethora of work that addresses ways to design constraints to keep the data consistent at all times. Functional dependencies and Denial Constraints are some examples~\cite{Xu13, Beskales14}. Implementing those constraints could be done either at the application level, i.e., write code to enforce those rules at the application, or at the DBMS level, i.e., write triggers that call a SQL procedure. The latter has the advantage of enforcing the constraints for any number of applications interacting with the database.

\item  \textbf{{Testing various hypotheses on the data:}}
It is estimated that designing data processing pipelines takes hundreds of iterations on average~\cite{mistique}. As the data undergoes a myriad of transformations over time, it is crucial to be able to quickly test hypotheses on the data to make sure the pipeline is up-to-date (e.g., new training data available, new updates to base tables). 

Workflow management systems can help design and test data pipelines as the data and the organizations specifications (e.g., new department) evolve. Data Civilizer is one example of such a tool~\cite{dc2}.

\item  \textbf{{Data version control:}} Typically, data is handled by different actors in an organization (different departments, etc.). It is important to keep track of “who did what” to the data over time. This makes it easy to ask the right people about versions of the data (instead of asking everyone in the organization). This is especially important when dirty data starts to appear as a result of a recent event (e.g., someone has inadvertently made a ``bad'' data update). An example system design is outlined in~\cite{elkindihilda}.

\item  \textbf{{Data discovery:}} Most organizations have a data warehouse or a data lake where they dump historical data for future analysis. As data analytics becomes a central part in modern organizations, it is important to make the job of “data scientists” as productive as possible. A well-cited figure states that data scientists spend over 80\% of their time preparing the data~\cite{dc2}. This leaves them with very little time to perform their analytical tasks. One key component of data preparation is Data Discovery. The premise of data discovery is: Given a set of tables (e.g., data lake), we would like to extract tables/records of interest to our task. 

\end{itemize}

\subsection{Data Privacy}

\begin{itemize}

\item  \textbf{Database Technologies for Privacy:}Protecting diverse datasets, their aggregates, and subproducts may require encrypted databases such as those presented in \cite{gadepally2015computing,popa2011cryptdb,poddararx,pappas2014blind}. Essentially, these systems attempt to enforce organizational policies on data confidentiality, integrity and/or availability of data throughout the data lifecyle. An overview of the state-of-the-art in encrypted databases is given in \cite{fuller2017sok}.

\item  \textbf{Data Ownership:} Limit data ownership and access to those with a need to know. While there are algorithmic solutions for such policy checking, these are often very difficult to implement and have major issues when something goes wrong. There are  new techniques proposed that may simplify the implementation of complex  access control policies such as query based access control \cite{shay2019don}; however, these approaches are still relatively new and in the research phase.
    
\item  \textbf{{Inadvertent Releases Are Very Possible:}} Unfortunately, ensuring data privacy is an enduring problem for which there are no easy solutions. In general, organizations can improve by minimizing data collection whenever possible, and by following good storage practices of sensitive data (such as storing data in encrypted form, and not allowing wide copies of sensitive files). These measures will help minimize the frequency of the most egregious "data spills", such as releasing an entire database of private financial information.
    
    However, even cautious organizations can make errors. A famous example is connected to the Netflix Prize contest of the mid-2000s. The Netflix Prize was a machine learning contest, in which Netflix challenged research teams to predict customers' movie ratings more accurately than any other team. As part of the contest, Netflix released a database of customer movie ratings, with all customer information replaced by an opaque identifier. The data had four fields: {\tt user-identifier}, {\tt movie-title}, {\tt grade}, and {\tt date-of-grade}.
    
    After Netflix released this data, researchers were able to identify the real names of some of the customers in this database. They examined the public rating history of users on IMDB.com, which reveals actual usernames. By looking for pairs of users who had distinctive and overlapping movie histories on both sites, they were able to tie the Netflix user identifiers to IMDB user accounts. Some of one user's Netflix rating history contained sensitive movie titles that were not part of the user's IMDB rating history, and which arguably revealed information about the user's sexual orientation.  
    
    In short: Netflix was careful to act responsibly in its data release, but still ended up enabling the revelation of private information.  Technical solutions to this problem are currently being explored in the research world (differential privacy is one direction), but are generally not deployable today.
    
    The US Census Bureau is a potentially useful model to emulate. They maintain a large amount of personal information about many Americans, which most would find extremely intrusive if it were released. This data is used by researchers and policymakers. However, the census data is believed to have never been part of an unintentional privacy breach, despite its long history\footnote{An arguable exception: census data was used to support the internment of Japanese-Americans during the Second World War, at the direction of government officials~\cite{aratani_2018}.}
    
    In order for researchers to use much of the Census Bureau data, they must:
    \begin{itemize}
        \item Undergo a background check. 
        \item Physically visit one of 29 Federal Statistical Research Data Center access points around the country, where they use a government-provided terminal.
        \item Not carry into the access point any electronic means of copying the data.
    \end{itemize}
    
    In other words, the Census Bureau has avoided Netflix's strategy of deidentification plus wide distribution. They have instead simply limited access to the data. This likely has come at some cost to researchers, but does seem to have achieved the data privacy goal.

\item  \textbf{{Sensitivity of aggregated data:}} Integration of different non-sensitive datasets may lead to sensitive outputs. For example, consider a medical scenario in which a user is trying to integrate data from \textit{Table 1} and \textit{Table 2}. Suppose \textit{Table 1} consists of columns (PatientID, Patient Name, Patient Gender) and \textit{Table 2} consists of columns (PatientID, Patient Date of Birth, Patient Age). If a researcher is interested in looking at the distribution of gender and age within the datasets, they may attempt to do a ``join'' on Table 1 and Table 2 using the PatientID. Very often, combining Patient Name and Date of Birth can lead to sensitive Personally Identifiable Information (PII). However, if the user instead filtered their queries to only select the pieces of data relevant from each table, the aggregated query may not lead to PII information. Similar cases exist within Department of Defense (DoD) and enterprise datasets. There are a couple of approaches to mitigate this risk. In one direction, you could run all services on a sensitive network. Thus, any data products will already be on a system or network approved for the aggregate (e.g., the Census Bureau example above). While this may seem like the easiest path forward for developers of the service, from a user perspective, this is not convenient (or feasible). Another approach may be to leverage encrypted databases ~\cite{kepner2014computing,hacigumucs2002executing,popa2011cryptdb,fuller2017sok} such that all data products are kept encrypted and can only be viewed if certain constraints are satisfied (e.g, a filter that ensures any data being decrypted doesn't contain sensitive aggregates). These tools, largely research efforts, are worth watching and may provide an interim solution. In the long term, we believe that these approaches in conjunction with automated tools built on knowledge graphs are a promising direction. Knowledge graphs would allow Information Security Officers (ISOs) of various organizations to relatively easily enforce various security policies. For example, a policy may be that {\tt Date of Birth} and {\tt Patient Name} should not be aggregated on a non PII-compliant system. With a knowledge graph representation, each record in the database is tied to an entity type and this policy is easy to enforce. In such a case, the data integration tool may be able to alert users that they are likely to have a policy violation and may even be able to help them construct a policy compliant query. A good overview of the challenges is given in \cite{uscertreport}.

\item  \textbf{{Experience with medical data access:}}
  We collaborate with two groups at the Massachusetts General Hospital to help them manage and clean their data. Managing access to this data has to be done carefully since it is sensitive data. As collaborators, we (1)~ cannot copy the data to any machine and have to use the data only within the MGH servers; and (2)~the data we have is de-identified. We also review any code that reads this data before deploying it. Data breaches often happen because of vulnerabilities in the code (e.g., web server assumes all requests are benign).
  
\end{itemize}

\noindent \subsection{Infrastructure, Technology and Policy}

\begin{itemize}
  
    \item \noindent {\bf Data Lakes:} A common software deployment is a so-called "data lake": a single physical location where all members of an organization place datasets, often in raw formats. The Hadoop Distributed Filesystem is a common infrastructure for data lakes, sometimes with some amount of administrative tooling.
    Importantly, there is usually little or no semantic integration; schemas might not even exist or be declared for all the data, let alone be integrated.
    
    Data lakes can serve some useful roles, but users will likely be disappointed if they hope the data lake alone will enable them to query the entire organization's data. Physically colocating the data is a potentially useful step, but semantic integration among datasets is often a much more time-consuming and expensive hurdle. Further, in many cases, physical colocation might not even be necessary for semantic integration.
    
    One scenario in which a data lake can be useful is when the data being uploaded is quite homogeneous. For example, it might be useful to transmit all of the log data from a set of servers to a single location. Another is when an organization has a sophisticated data science team that is ready to build novel analyses, and simply lacks physical access to all of the organization's data.

\item \noindent {\bf Database vs. Files:} Databases and files can often be used together. Use databases if you need to quickly find particular data items (i.e., you are looking for a small number of records when compared to the entire dataset).  SQL systems are good for medium size datasets that require ACID (atomicity, consistency, isolation, and durability) guarantees. NoSQL systems are good for large datasets where ACID guarantees aren’t required ~\cite{gadepally2015d4m}. NewSQL systems are good when you have a need for scalability and ACID compliance~\cite{pavlo2016s}. Scanning files in the file system can be best when reading in a majority of a dataset. 
    
    \item \noindent {\bf Federated Data Access:} Leverage a scalable data management architecture that allows the addition of new technologies such as object stores, databases and file systems. It is unlikely that any single technology solution will scale or provide efficient access to all modalities of data and federation as an architectural principle is important \cite{tan2017enabling}. As an example, polystore systems (an example, BigDAWG is described in detail in ~\cite{gadepally2016bigdawg,gupta2016cross, chen2016bigdawg}).
    
    \item \noindent {\bf Talent:} Access to the right people. Need for data scientists, subject matter experts in addition to data users. 
    \item \noindent {\bf Technology Selection:} Have Top-Down technology selection. Involve management in technology selection process. Avoid products/technologies that have unknown/unreliable development team (e.g., teams for adversary nations; teams of individuals unlikely to continue maintaining software)
    \item \noindent {\bf Software Licensing:}  Be aware of software licensing: Certain software libraries and products have restrictive software licenses. This may limit the ability to share technology with other industry/academic partners.
    \item \noindent {\bf Software and Hardware:} Avoid software/hardware products with unknown user base or non-active developer base. Reevaluate software/hardware products when the developers are acquired by other entities.
    \item \noindent {\bf Incorporating new tools:} Maintain an acquisition and development environment conducive to incorporating new technology advances. Many of the challenges discussed in this document are active areas of research in the commercial and academic sector. Innovations are coming at a rapid pace and it is important that decision makes maintain an environment to acquire and deploy these technologies in an agile fashion.

\end{itemize}
\newpage

\section{Acknowledgements}

The authors wish to thank the following individuals for their support in developing this report: Siddharth Samsi, Jeremy Kepner, Albert Reuther, Mike Stonebraker, Sam Madden. In addition, the authors wish to thank the MIT SuperCloud team: Bill Arcand, Bill Bergeron, David Bestor, Chansup Byun, Michael Houle, Matthew Hubbell, Michael Jones, Anna Klein, Peter Michaleas, Lauren Milechin, Julie Mullen, Andrew Prout, Antonio Rosa, Albert Reuther, Charles Yee.

This material is based upon work supported by the National Science Foundation under Grant No. 1636788 and by the United States Air Force Research Laboratory under Cooperative Agreement Number FA8750-19-2-1000. The views and conclusions contained in this document are those of the authors and should not be interpreted as representing the official policies, either expressed or implied, of the United States Air Force or the U.S. Government. The U.S. Government is authorized to reproduce and distribute reprints for Government purposes notwithstanding any copyright notation herein.

\newpage
\bibliographystyle{apalike}
\bibliography{references}
\end{document}